\documentclass[aps,amssymb,amsmath,prl,twocolumn]{revtex4-1}
\usepackage[breaklinks=true]{hyperref}
\usepackage[pdftex]{graphicx}
\usepackage{bm}
\usepackage[caption=false]{subfig}

\usepackage{color}
\usepackage[normalem]{ulem}

\begin{document}
\title{Spatial Correlations and the Insulating Phase of the High-$T_c$ Cuprates: Insights from a Configuration-Interaction-Based Solver for Dynamical Mean Field Theory}
\author{Ara \surname{Go}}
\author{Andrew J. \surname{Millis}}
\affiliation{Department of Physics, Columbia University in the City of New York, New York, NY 10027}
\begin{abstract}
A recently proposed configuration-interaction based impurity solver is used in combination with the single-site and four-site cluster dynamical mean field approximations to investigate the three-band copper oxide model believed to describe the electronic structure of high transition temperature copper-oxide superconductors. Use of the configuration interaction solver enables verification of the convergence of results with respect to  the number of bath orbitals.  The spatial correlations included in the cluster approximation substantially shift the metal-insulator phase boundary relative to the prediction of the single-site approximation and increase  the predicted energy gap of the insulating phase by about 1 eV above the single-site  result.  Vertex corrections occurring in the four-site approximation act to dramatically increase the value of the optical conductivity near the gap edge, resulting in a better agreement with the data.  The calculations reveal two distinct correlated insulating states: the `magnetically correlated insulator', in which nontrivial intersite correlations play an essential role in stabilizing the insulating state, and the strongly correlated insulator, in which local physics suffices. Comparison of the calculations to the data place the cuprates in the magnetically correlated Mott insulator regime. 
\end{abstract}
\date{\today}

\maketitle
\renewcommand{\k}{\mathbf{k}}
\newcommand{\K}{\mathbf{K}}
\newcommand{\R}{\mathbf{R}}
\newcommand{\wn}{\omega_n}
\newcommand{\Tr}{\mathrm{Tr}}

\newcommand{\del}[1]{{\color{red}\sout{#1}}}
\newcommand{\cor}[2]{{\color{red}\sout{#1}} {\color{blue}#2}}
\newcommand{\new}[1]{{\color{blue}#1}}

Understanding ``strongly correlated'' electron physics \cite{Imada1998}  is one of the grand challenges of condensed matter theory.  
The layered copper-oxide  materials such as La$_{2-x}$Sr$_x$CuO$_4$ are central to this endeavor because they exhibit a range of unusual electronic properties  including both high transition temperature superconductivity and a correlation-driven  insulating phase. Indeed, the physics that causes the insulating behavior is  believed \cite{Anderson1987} also to give rise to other important correlated electron properties, in particular,  superconductivity.  A clear understanding of the physics of the insulating phase is therefore essential. A basic question in the field is  whether the local effects of  strong  correlations are sufficient to describe the important properties  \cite{Anderson1987,Zhang1988,Lee2007,Weber2010} or whether  intersite correlations are essential to the description of the observed properties \cite{Schrieffer1988,Abanov2001,Comanac2008,Wang2011}.   
In this paper we use a cluster implementation \cite{Maier2005} of dynamical mean field theory \cite{Georges1996} to address the issue of the physics of the insulating phase of the cuprates.  A crucial role in our work is played by the configuration interaction (CI) solver introduced by Zgid, Gull and Chan \cite{Zgid2011, Zgid2012}, which enables the computation of converged real-frequency single-particle and optical spectra for wide parameter ranges including both strong and weak interactions.   We find that the ``copper-oxygen model'' which is generally believed \cite{Zaanen1985,Emery1987,Varma97}  to represent the basic electronic physics of the cuprates has three distinct regimes of behavior: a metal, a charge transfer insulator and a magnetically correlated charge transfer insulator in which the insulating behavior is due to intersite correlators and not to the standard local Mott physics.  Comparison of our results to data locates the cuprates in the magnetically mediated insulator regime.

An appropriate `microscopic' Hamiltonian for the materials  is $H_{CT}=H_d+H_{rest}$  
\begin{eqnarray}
H_d&=&\sum_{k\sigma} \varepsilon_dd^\dagger_{k\sigma}d_{k\sigma}+U\sum_in_{d,i\uparrow}n_{d,i\downarrow}
\label{Hd}
\\
H_{rest}&=&\sum_{ka\sigma}t_{pd}^a(k)d^\dagger_{k\sigma}p^a_{k,\sigma}+H.c+ \sum_{kab\sigma}\varepsilon_k^{ab}p^{\dagger,a}_{k\sigma}p^b_{k\sigma}.
\label{Hrest}
\end{eqnarray}
where $k$ is a momentum in the two-dimensional Brillouin zone, $d^\dagger_{k\sigma}$ creates an electron of momentum $k$ in a Cu orbital and $p^{\dagger,a}_{k\sigma}$ creates an electron in one of the two in-plane oxygen $p_\sigma$-orbitals. The charge transfer parameter $\Delta$ is defined as the difference between the unrenormalized on-site copper energy $\varepsilon_d$ and the average on-site oxygen energy $\varepsilon_p=\frac{1}{2}\mathrm{Tr}_{kab}\varepsilon_k^{ab}$ as $\Delta=\varepsilon_p-\varepsilon_d$. 

The parameters of $H_{CT}$ may be derived e.g. from Wannier function fits to a band calculation; however the $d$ energy $\varepsilon_d$ must be renormalized by a ``double counting correction'' whose magnitude is not theoretically known \cite{Karolak10}. Previous work \cite{Wang2011} has shown that the behavior of the model does not depend on the details of the oxygen dispersion $\varepsilon_k^{ab}$ or on how the double counting is implemented. The only important variable is the $d$ occupancy $N_d=\langle d^\dagger_{i\sigma}d_{i\sigma}\rangle$, which of course depends on these variables in a complicated way. In this Letter we therefore adopt the most convenient model, $\varepsilon_k^{ab}=\varepsilon_p\delta_{ab}$, $t_{pd}^a(k)=2i \sin k_a$ and regard the double counting correction (i.e. the $p$-$d$ energy difference $\Delta$) as a parameter of the theory. 

We study $H_{CT}$ by using the single- and four-site versions of the dynamical cluster approximation implementation  of  dynamical mean field theory \cite{Maier2005} as applied to the three-band model e.g., by Macridin \textit{et al.} \cite{Macridin05}. Previous work on the Hubbard model revealed  large qualitative differences between the single-site and four-site cluster results \cite{Maier2005,Gull10_clustercompare} with larger clusters providing important differences of detail but not changing the qualitative picture \cite{Gull10_clustercompare}. Less work has been done on cluster approximations to  $H_{CT}$ although the validity of the one-band model has been considered \cite{Macridin05}, and an interesting studies of the dependence of superconducting properties on the apical oxygen distance has appeared \cite{Weber2010}. 

The central computational task in dynamical mean field theory  \cite{Georges1996} is  the solution of a ``quantum impurity model'', a $0$-space plus $1$-time dimensional quantum field theory or alternatively  a small number $N_c$ of interacting orbitals coupled to  a noninteracting bath. The existing methods of solution are not fully satisfactory.   Continuous-time quantum Monte-Carlo method\cite{Rubtsov2005,Werner2006,Gull2011} has proven effective for the  single-band Hubbard model at not too strong correlations \cite{Fuchs2011} and for multiorbital situations in the single-site approximation \cite{Werner2006,Haule07} but scales very poorly with system size in situations involving orbital degeneracy, becomes prohibitively expensive for strong correlations and suffers a severe sign problem in situations with more than one orbital and low point symmetry \cite{Gull2011}. Also it is formulated in imaginary time and an ill-controlled analytical continuation process is required to obtain the real-frequency information required for spectral functions. The numerical renormalization group \cite{Bulla2008} and the density matrix renormalization group \cite{Hallberg2006}  have been effective in special situations (for example determining the precise low-frequency spectral properties of the single-orbital Hubbard model in the single-site approximation) but have proven difficult to apply generally. The exact diagonalization method of Caffarel and Krauth \cite{Caffarel1994} and improved by Capone \cite{Capone2007} approximates the quantum field theory as a finite-size Hamiltonian which is diagonalized by using e.g., Lanczos methods, and although interesting studies have appeared \cite{Liebsch2008,Liebsch2009,Liebsch2012,Civelli2008},  is limited by the number of  sites available. 

In this Letter we use our implementation of a new method, the configuration interaction approach of Zgid and Chan \cite{Zgid2012}, to study the metal-insulator phase diagram, electron spectral function and optical conductivity of a fundamental model of the high transition temperature CuO$_2$-based superconducting materials.  We use a zero temperature implementation. A related CI implementation has recently been used by Lin and Demkov \cite{Lin2013} to study defect and other properties of $\mathrm{SrTiO}_3$.  The CI method is a variant of ED in which the full Hilbert space is not treated; rather, the diagonalization is performed in a variationally chosen subspace, allowing  larger problems to be attacked. The details of our implementation will be given elsewhere \cite{Go2014}. Here we note that  the ground state is found by minimizing the Hamiltonian in a subspace consisting of number $N_\mathrm{ref}$ of reference states  plus all possible states containing up to $P$ particle-hole pair excitations above the reference states. The key to the method is that $P$ is small.  We find that, in general, choosing up to two particle-hole pairs for each spin direction (this is a subset of all possible $P=4$ states) suffices, and that for moderate interactions $U\lesssim 12\mathrm{eV}$ simply restricting to $P=2$ suffices.

\begin{figure}[b]
	\includegraphics[type=pdf,ext=.pdf,read=.pdf,width=0.99\linewidth]{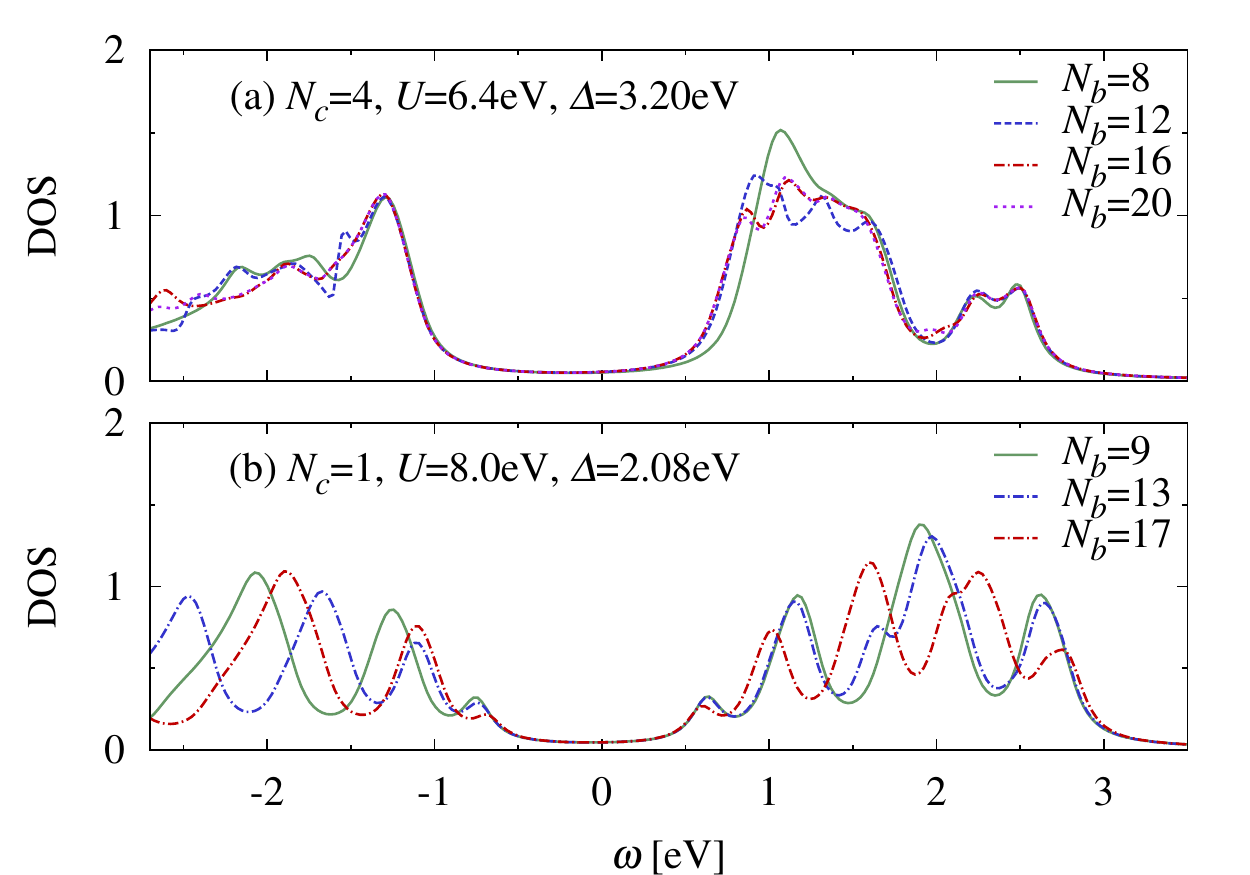}
  \caption{\label{fig.bath}%
  (color online) Density of states from (a) four-site and (b) single-site DMFT with various values of $N_b$.
  A small broadening factor $\eta=0.10\mathrm{eV}$ is used.
  }
\end{figure}

The reference states are obtained in practice as follows. We define the natural orbital basis as the eigenstates of the single-particle density matrix of the ground state $\hat{\rho}=|\psi \rangle \langle \psi |$. We choose as active orbitals  the $2N_c$ single-particle states of the natural orbital basis with ground-state occupancy closest to $1/2$. The other orbitals are found to be a very good approximation to have occupancy 1 or 0. As also noted by Lin and Demkov \cite{Lin2013}, this limited number of partially filled orbitals is crucial to the success of the method. The reference states are then  defined as all many-body states which may be formed from the $2N_c$ active orbitals with appropriate conserved quantities (these are particle number $N_\mathrm{act}=N_c$ and, in the four-site calculations,  spin $S_z=0$) with the other orbitals remaining filled or empty.    Since the reference states and the ground state depend on each other, the whole procedure is iterated until self-consistency is reached. 

In this scheme the number $N_\mathrm{ref}$ of reference states is equal to the number of states in the largest $S_z=0$ sector of the impurity subspace, i.e., $N_\mathrm{ref}=( {}_{N_c} C_{N_c/2} )^2$ where $_nC_m=n!/[m!(n-m)!]$, so that while the computational complexity grows exponentially in the size of the impurity Hilbert space, it scales only as a power of the number of bath orbitals and the reduced   dimension of the Hilbert space in CI means the prefactor is smaller. Studies of up to  $N_b=20$ are possible for a four-site cluster without  parallelization for distributed memory system; larger systems should be accessible when the algorithm is optimized.  

Figure~\ref{fig.bath} shows the density of states (DOS)  $\rho$ obtained from converged DMFT solutions for different number of bath states $N_b$.
As is standard in ED calculations, a small broadening factor $\eta=0.10~\mathrm{eV}$ is introduced
and $\rho$ is defined in terms the trace of the branch cut discontinuity of the local Green function: $\rho(\omega)=\mathrm{Tr}\left[\mathbf{G}_{\mathrm{loc}}(\omega-i\eta)-\mathbf{G}_{\mathrm{loc}}(\omega+i\eta)\right]/(2\pi i)$. $\mathbf{G}_{\mathrm{loc}}$ is the $12\times 12$ matrix  (4 momentum space tiles and three orbitals per tile) defined in Eq. (7) of the Supplementary Material.   Consistent with previous work \cite{Luca2009}, our single-site calculations (lower panel) show that $N_b=9$ is sufficient to describe the behavior at gap edge. For the four-site case we verify that $N_b=8$ produces qualitatively correct results, but leads to errors in the spectral gap of $\sim 0.2~\mathrm{eV}$, while for $N_b=12$ the spectral gap is quantitatively converged. In the rest of the Letter we use $(N_c, N_b)=(1,9)$ and $(4,12)$ unless otherwise mentioned.

\begin{figure}[tbp]
	\includegraphics[type=pdf,ext=.pdf,read=.pdf,width=0.95\linewidth]{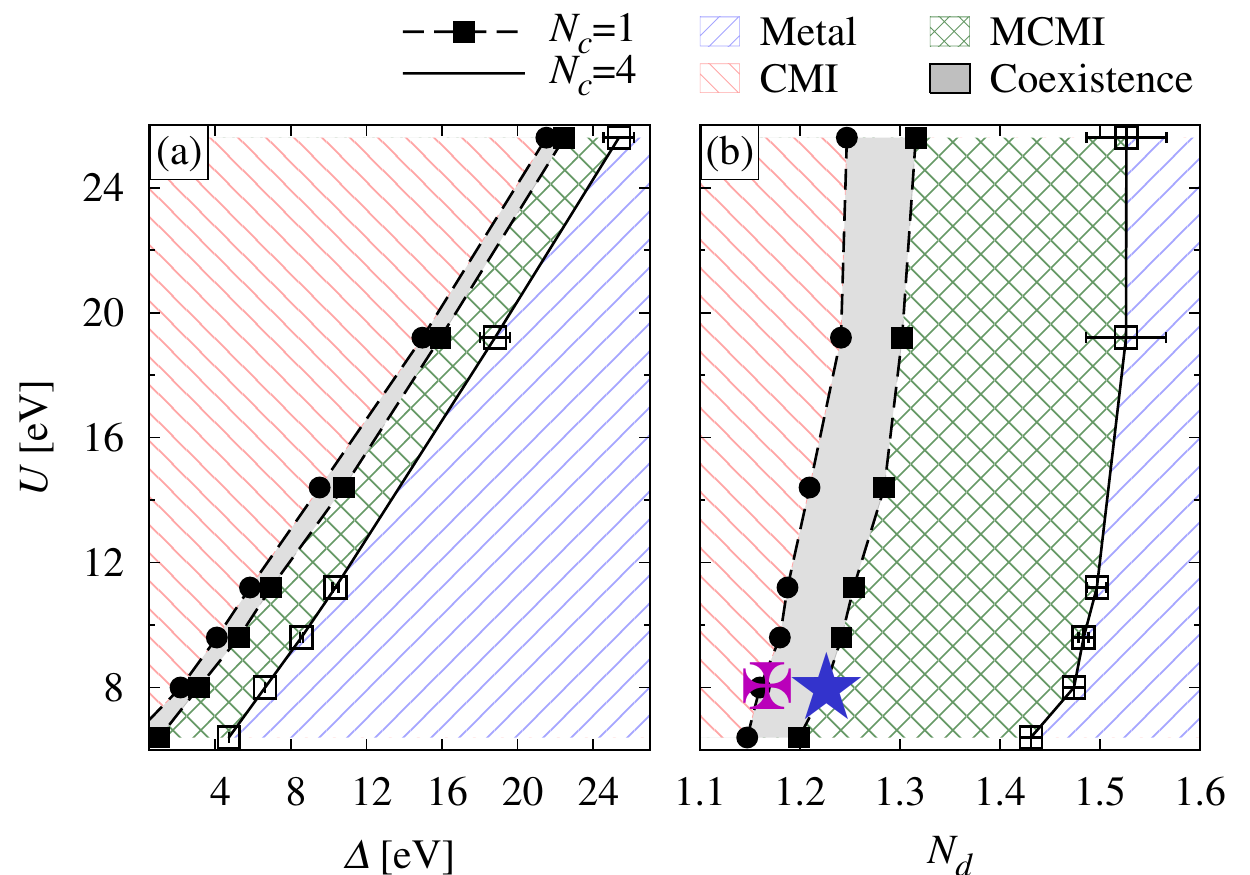}
  \caption{\label{fig.phase}%
  (color online) Metal-insulator transition phase diagram in plane of interaction strength $U$ and $p$-$d$ energy splitting $\Delta$ (panel  (a)) and $d$-occupancy $N_d$ (panel (b)) from single-site ($N_c=1$) and four-site ($N_c=4$) dynamical mean field approximation. The error bars reflect uncertainties arising from restricting to  $P=2$ particle-hole pairs at large U; where not shown they are smaller than the size of the points.   The region shaded with lines slanting up and to the left is the conventional Mott insulator (CMI) region, where  insulating  behavior is found even for $N_c=1$; the region shaded with lines slanting up and to the right indicate regions that are metallic in the single-site approximation and the cross-hatched region is the magnetically correlated Mott insulator (MCMI), which is insulating for $N_c=4$ but not $N_c=1$.  The shaded area is the coexistence region of the metal-insulator transition in the single-site approximation. The magenta cross and the blue star  denote parameter values that yield gaps comparable to the experimental values for $N_c=1$ and $4$ respectively.
  }
\end{figure}

In the left panel of Fig.~\ref{fig.phase} we present the metal-insulator phase diagram obtained by single-site and cluster DMFT method in the plane of interaction strength and charge transfer energy at total filling $n=5$ (one hole per Cu). To determine the nature (metallic vs insulating) of the state we examine the low frequency behavior of the self-energy. For an insulator, the self-energy has a pole near the chemical potential, while for a metal the self-energy is smooth. This criterion is less sensitive to finite-bath size errors than is a direct examination of the density of states. The single-site approximation contains the physics of conventional Mott/charge transfer insulators, while the cluster approximation additionally includes the effects of intersite correlations. Comparison of the two allows us to distinguish different types of insulators.  

The  single-site results are consistent with previous work \cite{Luca2009}. The transition is first-order; the coexistence region is shaded in Fig.~\ref{fig.phase}. The size of this coexistence region is robust against increasing $N_b$.  The four-site approximation widens the insulating regime, shifting the phase boundary in the $U-\Delta$ plane to the right by about $\delta \Delta=3~\mathrm{eV}$. The shift indicates that (as also found in the Hubbard model \cite{Maier2005,Park2008,Gull08,Gull10_clustercompare}) intersite magnetic correlations present in the $N_c=4$ but not the $N_c=1$ calculation play a crucial role in stabilizing the insulating state. We  designate the region which is insulating only if intersite correlations are included as the magnetically correlated Mott insulator (MCMI) and the region which is insulating even in the single-site approximation as the conventional Mott insulator (CMI).  As in the $4$-site approximation to the  Hubbard model \cite{Park2008,Gull08}, the transition in the four-site approximation to the $p$-$d$ model is found to be weakly first-order, with a small coexistence region; however the size of the coexistence region shrinks as the number of bath states is increased (not shown) and the actual size of the coexistence region for this model is not established by the results we have.  

Previous work \cite{Wang2012} has shown that it is useful to consider the physics as a function of the $d$-occupancy $N_d$. In $p$-$d$ models of the kind studied here, results expressed in terms of $N_d$ are insensitive to such details of the band structure as the oxygen-oxygen hopping. The right panel of Fig.~\ref{fig.phase}  shows the metal-insulator phase diagram in the  plane of interaction strength and $d$-occupancy. Inclusion of intersite correlations shifts the phase boundary by about $0.25$ in $N_d$, with the shift  being independent of  $U$. The ability of the CI method to attain larger $U$ allows us to see that the phase boundary in the $U-N_d$ plane becomes vertical only for very large $U\sim 16~\mathrm{eV}$, while physically relevant values for the copper-oxide materials are $\sim$ 8--10~eV \cite{Imada1998}.

\begin{figure}[tbp]
	\includegraphics[type=pdf,ext=.pdf,read=.pdf,width=0.99\linewidth]{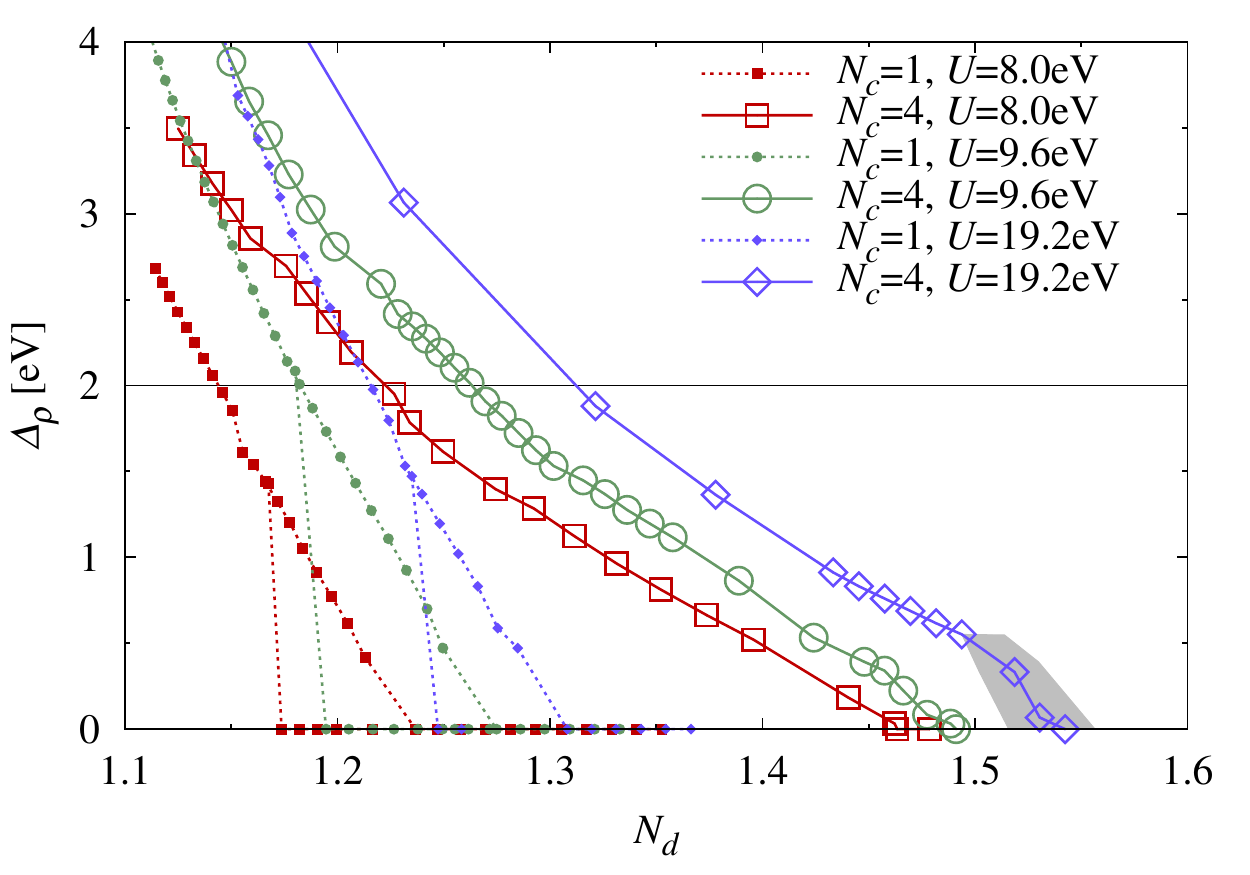}
  \caption{\label{fig.class}
  (color online) Spectral gap as a function of $d$-occupancy for various values of $U$ with $(N_c,N_b)=(1,9)$ and $(4,12)$. For $N_c=1$, two distinct solutions depending on initial conditions are shown in the intermediate region. The black horizontal line indicates the charge transfer gap $\Delta_\rho=2~\mathrm{eV}$ characteristic of the parent compounds of the  high-$T_c$ cuprates. The size of the shaded region reflects the uncertainties arising from the $P=2$ approximation at large $U$.}
  
\end{figure}

In Fig.~\ref{fig.class} we present the excitation gap determined from the calculated self-energy and the quasiparticle equation as discussed in Ref.~\cite{Wang09gap}.  In the single-site approximation, at fixed $U$ the gap magnitude in the insulating solution decreases linearly as $N_d$ increases. This smooth behavior indicates that there is only one kind of insulating state in the single-site approximation. We identify this as the CMI phase. As extensively discussed \cite{Georges1996}, the gap in the CMI phase can be smoothly decreased to zero but this transition is preempted by a first-order transition to a metallic phase. In the four-site approximation, two regimes are evident: a small $N_d$ strongly correlated regime where the gap vs $N_d$ curve is very similar to the single-site approximation (albeit with an enhanced gap) and a larger $N_d$ regime where the slope of the $\Delta-N_d$ curve changes. We identify this regime as the MCMI. The crossover between the two regimes occurs at the point at which the gap closes in the single-site approximation.   

The optically determined \cite{Uchida1991} charge transfer gap of about $2~\mathrm{eV}$ is shown as a horizontal line. For $U$-values in the generally accepted range  $U\sim 8$--10~eV we see that an $N_d\sim1.14$--1.20 is required to reproduce the gap  in the single-site approximation; in the four-site approximation a larger $N_d\sim 1.22$--1.28 is needed to reproduce the gap. The $N_d$ values needed to reproduce the insulating gap for $U=8~\mathrm{eV}$ are marked in Fig.~\ref{fig.phase} by the symbols `$\maltese$' (for the single-site case) and `$\bigstar$' (in the four-site case). The $N_d$ needed to fit the observed excitation gap in the four-site approximation is in the coexistence region of the single-site approximation. Both in the single-site and four-site cases the $N_d$ values required to account for the observed gap are substantially smaller than the density functional prediction $N_d\sim 1.4$ \cite{Wang2012}, suggesting that density functional theory overestimates the Cu-O covalence. An analysis of nuclear magnetic resonance data \cite{Haase2004}  suggests an $N_d\sim1.22$ consistent with the $N_c=4$ calculation.   

\begin{figure}[tbp]
	\includegraphics[type=pdf,ext=.pdf,read=.pdf,width=0.99\linewidth]{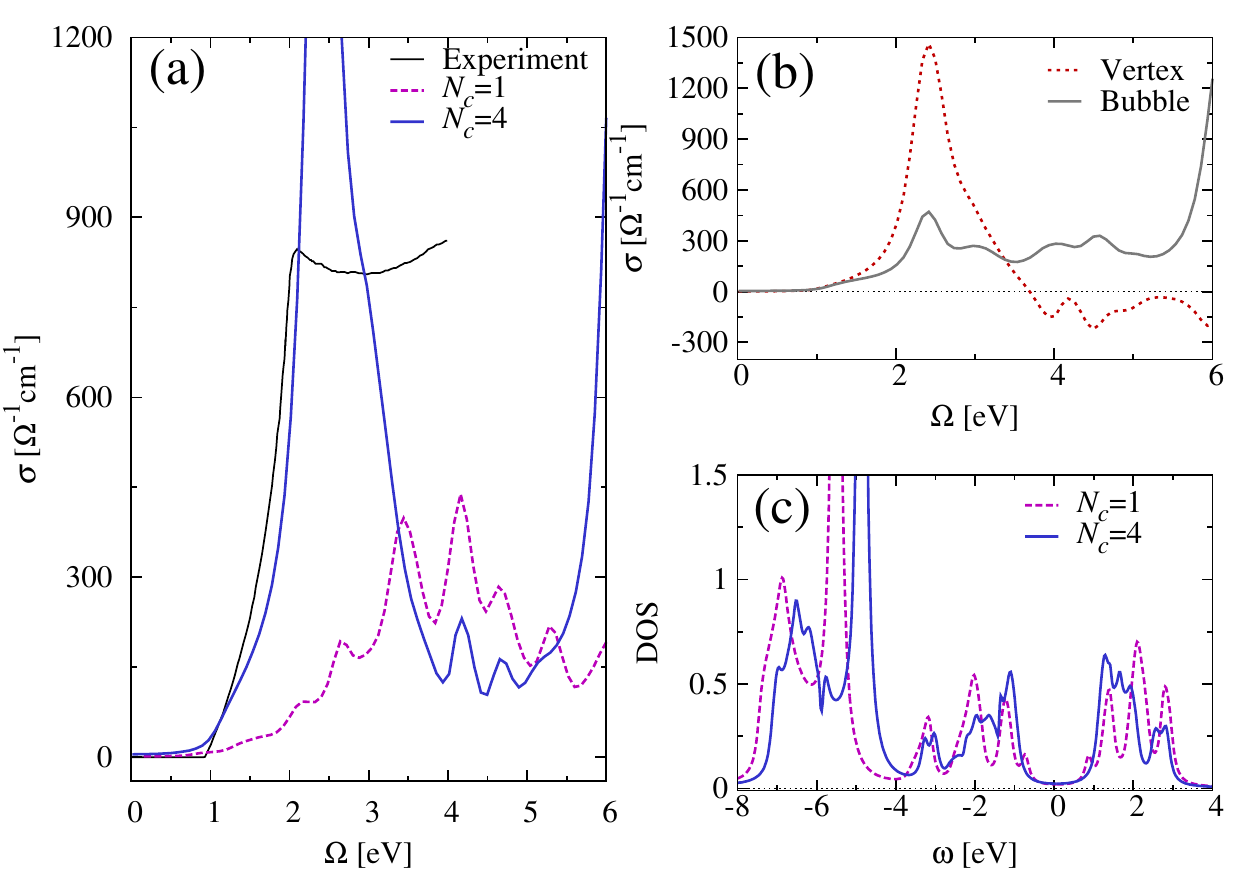}
  \caption{\label{fig.opt_u5}%
  (color online)
  (a) Optical conductivity from the single- and four-site DMFT with parameters fixed to yield gap size $\Delta_\rho\sim2~\mathrm{eV}$ at correlation strength $U=8.0~\mathrm{eV}$ while the four-site calculation leads to a rapid rise in the conductivity at the gap edge, in agreement with experiment.  
  (b) Vertex and bubble contribution to the optical conductivity for four-site DMFT.
  (c) Density of states with the same parameters in (a).
   }
\end{figure}

Figure \ref{fig.opt_u5} presents the optical conductivity (a, b) and DOS (c) obtained for $U=8~\mathrm{eV}$ with $\Delta$ values chosen to reproduce the observed $\sim 2$~eV gap. In the four-site optical conductivity calculation,  vertex corrections are incorporated following the method presented in Ref.~\cite{Lin2009}. We see that the single-site calculation predicts a very small value for the optical conductivity at frequencies not too far above the upper gap edge, while the four-site calculation  yields a much larger conductivity for frequencies near the gap edge, in a better agreement with data \cite{Uchida1991}.  The physics is that the gap is indirect and vertex corrections (not present in the $N_c=1$ calculation) activate gap edge transitions by allowing for a multiparticle transition in which an excitation of momentum $\mathbf{Q} \approx (\pi,\pi)$ is emitted. The vertex corrections are present for all values of the correlation strength but are most important in the MCMI regime (see the Supplementary Material). The large enhancement of the gap edge conductivity relative to experiment is an artifact of the $N_c=4$ approximation, which concentrates the vertex corrections at the boundaries between momentum space tiles. The integrated spectral weight, which is more robust to details of methodology is in good agreement with data (see supplementary material).

In summary, this Letter introduces an implementation of the Zgid-Chan CI solver \cite{Zgid2011} which allows us to obtain converged real-frequency results for single-particle and conductivity spectra of the charge-transfer model generally agreed to represent the physics of the high-$T_c$ cuprate superconductors, for a wide range of previously inaccessible parameters. Our results enable us to distinguish two types of insulating phase, the conventional Mott insulator and the magnetically correlated Mott insulator and  comparison of our calculations to experiment place the materials in the MCMI region of the phase diagram, supporting previous suggestions \cite{Comanac2008,Wang2011,Wang2012} that intersite correlations play an essential role in the physics of the high-$T_c$ cuprates. A subsequent paper will investigate the different doping dependences of the two phases. Our work also resolves a previously noted \cite{Wang2011} discrepancy between theory and optical conductivity data. Finally, we confirm previous indications \cite{Wang2012} that single-site dynamical mean field theory provides a quantitatively poor approximation to basic properties of the two-dimensional charge-transfer model and  that density functional band theory overestimates the $p$-$d$ hybridization and should not be used as a guide for placing materials on the metal-charge transfer insulator phase boundary.   Our work validates the CI method as a robust and powerful approach for investigating the physics of correlated electron materials. As a direction for future work we note that the ability of the CI method to treat a much larger number of bath orbitals than is possible in conventional ED solvers indicates that the method will be useful in treating  the  nondiagonal hybridization functions  arising in low-symmetry situations, where severe sign problems limit the applicability of quantum Monte Carlo methods \cite{Gull2011} and difficulties with bath fitting prevent the application of conventional ED methods. Spin-orbit coupled situations and  cluster DMFT descriptions of systems with several partially occupied correlated orbitals may now be theoretically accessible. 

This work was supported by the U.S. Department of Energy under Grants No. DOE FG02-04ER46169 and No. DE-SC0006613.

\clearpage

\section{Supplementary Material}\label{sec:methods}

\subsection{Bath fitting}
\begin{figure}[btph]
	\subfloat[]{\includegraphics[type=pdf,ext=.pdf,read=.pdf,width=0.49\linewidth]{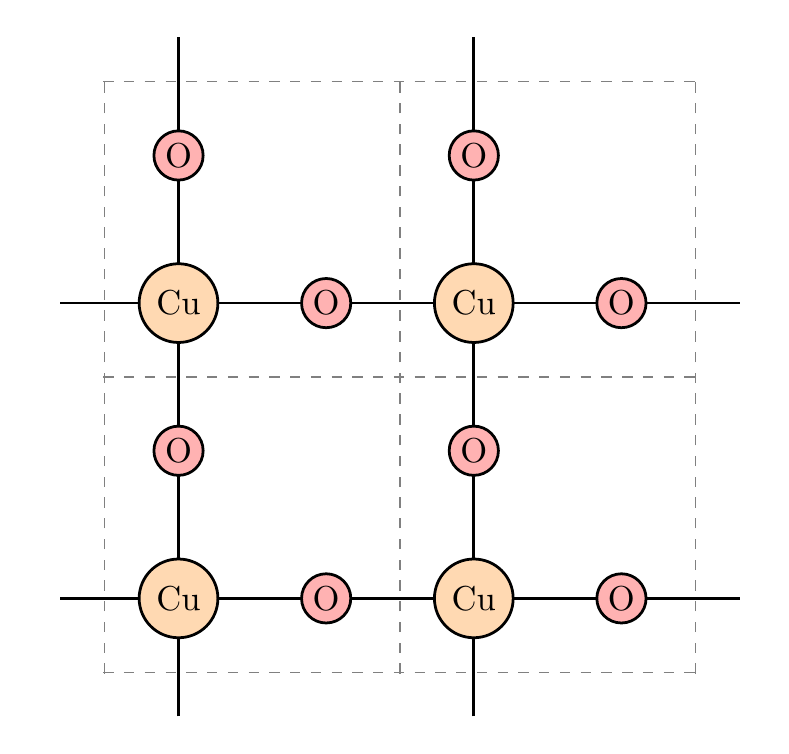}}
	\subfloat[]{\includegraphics[type=pdf,ext=.pdf,read=.pdf,width=0.49\linewidth]{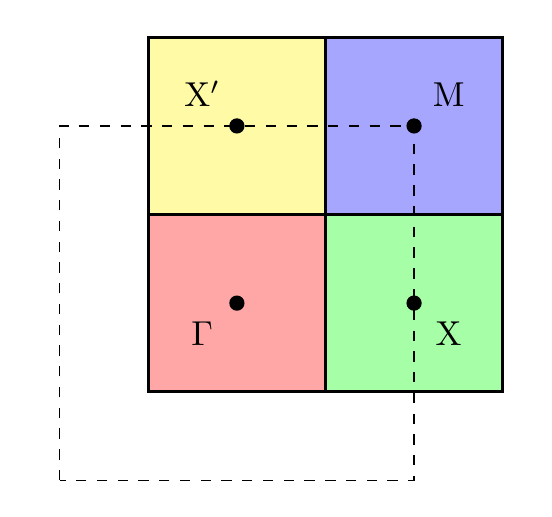}}
  \caption{\label{fig.model}%
  (color online) (a) Lattice structure of the three-band model and (b) four tiles in the first Brillouin zone.
  }
\end{figure}

The first step to build an appropriate self-consistent equation for dynamical mean field theory (DMFT) with exact diagonalization (ED) solver is finding an effective impurity (cluster) Hamiltonian which is consistent with the lattice Hamiltonian.
The noninteracting model Hamiltonian in a unit cell of Fig.~\ref{fig.model}(a) is represented as $3\times 3$ matrix,
\begin{align}
	\hat{H}_0(\k) = 
	\begin{pmatrix}
		\varepsilon_d 	& 2i \sin(k_x/2)	& 2i \sin(k_y/2)	\\
		-2i \sin(k_x/2)	& \varepsilon_p		& 0		\\
		-2i \sin(k_y/2)	& 0			& \varepsilon_p	
	\end{pmatrix},
\end{align}
where $\varepsilon_p = \varepsilon_d + \Delta$ and the circumflex down symbol `$\hat{~}$' denotes $3\times 3$ matrix.
We assume that only the copper $d$-orbital has a non-negligible local interaction and therefore the self-energy is nonzero only for the $d$-orbital.
In the dynamic cluster approximation (DCA), the self-enbergy is piecewise constant in momentum space. Dividing momentum space into tiles in which the self-energy is constant, the lattice Green function of each tile is obtained as
\begin{align}
	\hat{G}_\K(\omega) = \sum_{\k \in \K} \big[ \omega \cdot \hat{1} - \hat{H}_0(\k) - \hat{\Sigma}_\K(\omega) \big]^{-1},
\end{align}
with the self-energy
\begin{align}
	\hat{\Sigma}_\K(\omega) = 
	\begin{pmatrix}
		\Sigma_\K(\omega) & 0 & 0 \\
		0 & 0 & 0 \\
		0 & 0 & 0
	\end{pmatrix},
\end{align}
and the unit matrix $\hat{1}$.
In the single-site approximation, the tile in momentum space is the first Brillouin zone itself and the Green function is $3 \times 3$ matrix
while that of the cluster DMFT is $3N_c \times 3N_c$ matrix.
To incorporate the lattice symmetry, it is convenient to transform the Green function to the site-basis.
The Green function on the momentum basis is diagonal in DCA approach and it can be transformed easily.
In the four-site approximation, for example, the Green function is given by applying a $12\times 12$ matrix (each element of following $4\times 4$ matrix is a $3\times 3$ matrix),
\begin{align}
	\tilde{T} = 
	\frac{1}{2}
	&\begin{pmatrix}
	\begin{array}{rrrr}
		\hat{1}	& \hat{1}	& \hat{1}	& \hat{1}	\\
		\hat{1}	& -\hat{1}	& \hat{1}	& -\hat{1}	\\
		\hat{1}	& \hat{1}	& -\hat{1}	& -\hat{1}	\\
		\hat{1}	& -\hat{1}	& -\hat{1}	& \hat{1}
	\end{array}
	\end{pmatrix},\\
	\tilde{G}(\omega) =
	\tilde{T}
	&\begin{pmatrix}
		\hat{G}_\Gamma & 0 & 0 & 0\\
		0 & \hat{G}_\mathrm{X} & 0 & 0 \\
		0 & 0 & \hat{G}_{{\mathrm{X}^\prime}} & 0 \\
		0 & 0 & 0 & \hat{G}_\mathrm{M}
	\end{pmatrix}
	\tilde{T}^{-1}.
\end{align}
Each tile covers the same area in momentum space as shown in Fig.~\ref{fig.model}(b).
The Weiss field to construct the effective Hamiltonian for DMFT is defined as an $N_c \times N_c$ matrix and the elements are obtained by collecting $d$-orbital parts in which the interactions are not neglected,
\begin{align}
	[\bar{\mathcal{G}}_{\mathrm{target}}^{-1}(\omega)]_{\mu\nu} =  [\tilde{G}^{-1}(\omega) + \tilde{\Sigma}(\omega)]_{3\mu,3\nu},
	\label{eq.Weiss_new}
\end{align}
where $\mu,\nu=1,\cdots,4$ label cluster indices in the four-site cluster DMFT.
On the other hand, the Weiss field is also written in terms of impurity parameters,
\begin{align}
	[\bar{\mathcal{G}}_0^{-1}(\omega)]_{\mu\nu} =  \omega \delta_{\mu\nu} - \bar{t}_{\mu\nu} - \sum_{l=1}^{N_b} \frac{V^{}_\mu V^{*}_\nu}{\omega - \epsilon_l},
\end{align}
where the effective impurity Hamiltonian reads
\begin{align}
	{H}_\mathrm{eff} &=
	\sum_{\mu\nu\sigma} \bar{t}_{\mu\nu} d^\dagger_{\mu\sigma} d^{}_{\nu\sigma}
	+ U \sum_{\mu} n^{}_{\mu\uparrow} n^{}_{\mu\downarrow} \nonumber\\
	&+ \sum_{l\sigma} \epsilon_l a^\dagger_{l\sigma} a^{}_{l\sigma} + \sum_{\mu l \sigma} (V^{}_{\mu l} a^{\dagger}_{l\sigma} d^{}_{\mu\sigma} + \mathrm{H.c.}) 
\end{align}
where $\bar{t}_{\mu\nu} = [\sum_\k \tilde{T} \tilde{H}_0(\k) \tilde{T}^{-1}]_{3\mu,3\nu}$ is set to satisfy the self-consistency.
The other parameters $\{\epsilon_l, V_{\mu l}\}$ are obtained from bath fitting procedure.

\begin{figure}[tbp]
	\includegraphics[type=pdf,ext=.pdf,read=.pdf,width=0.99\linewidth]{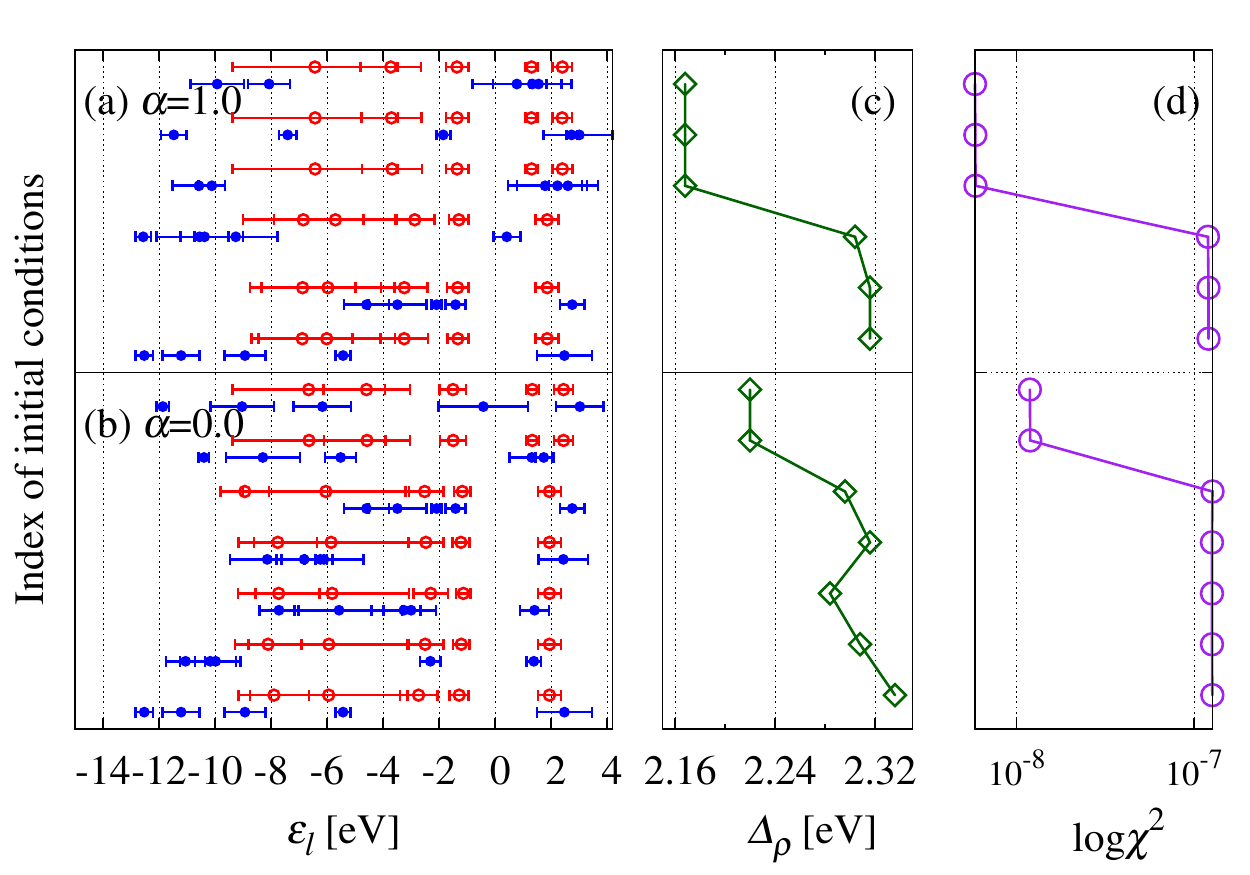}
  \caption{\label{fig.init_nb5}%
  (color online) Converged bath parameters from various random initial conditions with (a) $\alpha=1.0$ and (b) $\alpha=0.0$ for $N_b=5$, $U=8.0$~eV and $\Delta=1.12$~eV.
  Each set of filled blue and empty red circles are located at the bath energy level $\epsilon_l$ of the initial condition and the converged solution, respectively.
  The length of the errorbar is equal to the corresponding hybridization strength, $V_l$.
  The spectral gap and the distance function from each converged solution are shown in (c) and (d) with the same vertical location, respectively.
  The final solutions are roughly categorized by number of positive energy bath orbitals.
  }
\end{figure}

\begin{figure}[tbp]
	\includegraphics[type=pdf,ext=.pdf,read=.pdf,width=0.99\linewidth]{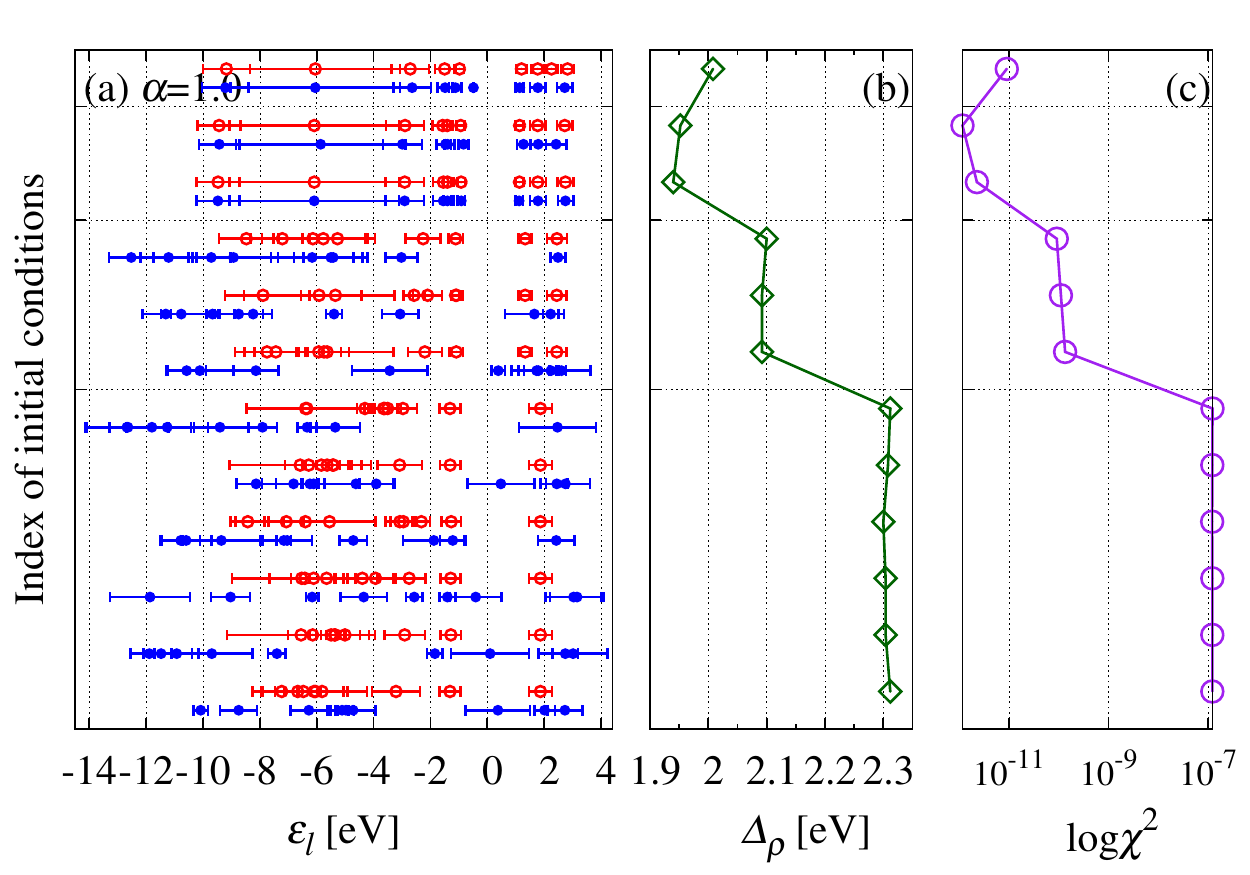}\\
	\includegraphics[type=pdf,ext=.pdf,read=.pdf,width=0.99\linewidth]{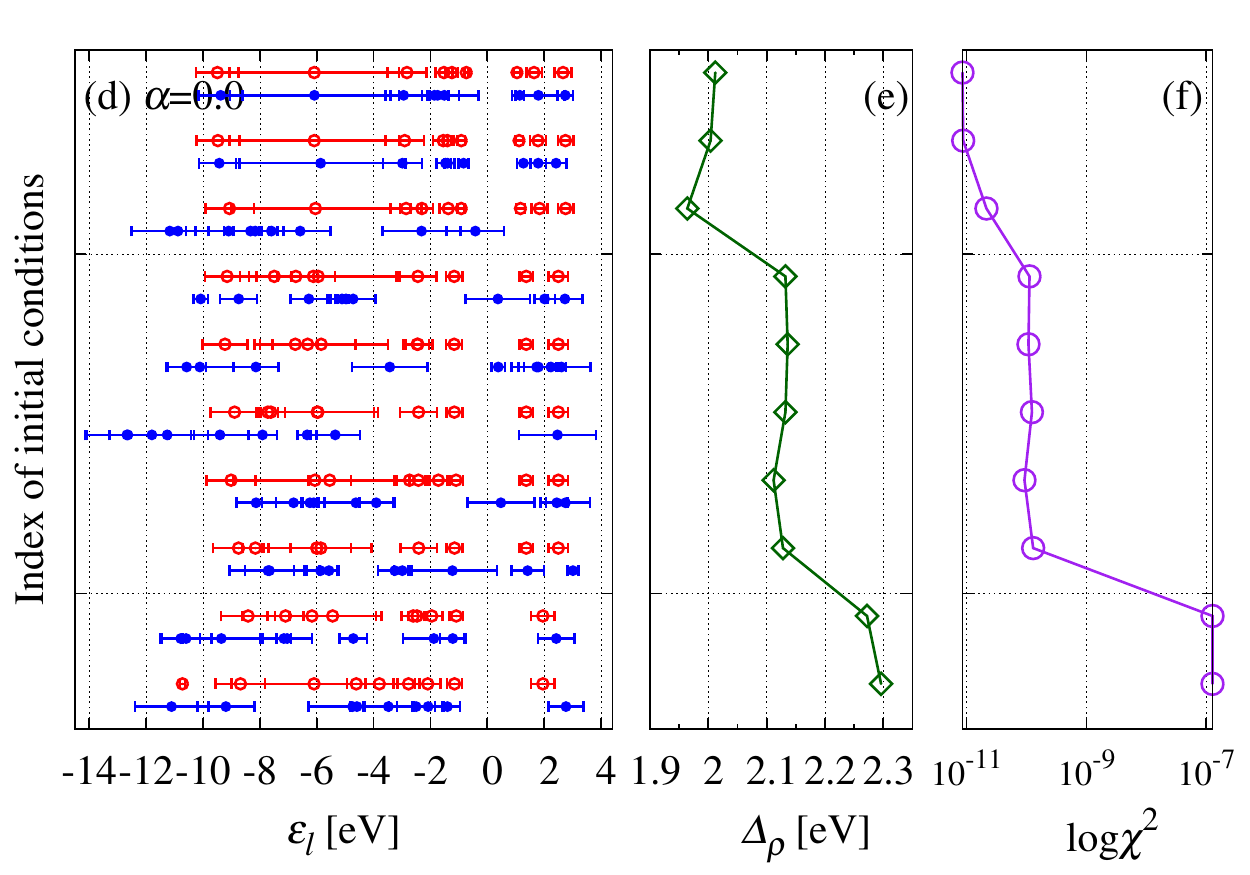}
  \caption{\label{fig.init_nb9}%
  (color online) Converged bath parameters from various random initial conditions for the same parameters with Fig.~\ref{fig.init_nb5} except $N_b=9$.
  }
\end{figure}

In practice, the DMFT+ED procedure starts from solving the impurity Hamiltonian with an appropriate initial bath parameters $\{\epsilon_l, V_{\mu l}\}$.
The impurity solver enables us to compute the self-energy and we extract the corresponding Weiss field by using Eq.~(\ref{eq.Weiss_new}).
(A disadvantage of ED as an impurity solver appears at this point: the number of bath orbitals $N_b$ is limited due to exponentially increasing computational costs with respect to the system size.)
Once the target Weiss field is obtained, we find the best fit to reproduce it by minimizing a distance function defined as
\begin{align}
	\chi^2 = \frac{1}{N_\mathrm{max}+1} &\sum_{n=0}^{N_\mathrm{max}} W(\omega_n) \nonumber\\
	&\times \sum_{\mu\nu} | [\bar{\mathcal{G}}^{-1}_\mathrm{target}(i\omega_n) - \bar{\mathcal{G}}^{-1}_0(i\omega_n) ]_{\mu\nu} |^2
\end{align}
where the Matsubara frequencies $\omega_n = (2n+1)\pi/\beta$ is given by the fictitious inverse temperature $\beta$, $N_\mathrm{max}$ is the upper limit of the summation,  and $W(\omega_n)$ is a frequency-dependent weight function.
The new bath parameters obtained by the conjugate gradient method are sent to the impurity solver to calculate a new self-energy and the self-energy produces a new Weiss field via a new local Green function.
The procedure continues until the convergence is reached.
Since $N_b$ is a finite number, there is no unique way to define the best fit when we extract the new bath parameters
and the converged DMFT solution more or less depends on the definition of the distance function $\chi^2$ and the initial condition.

Here we examine the weight function $W(\omega_n)=1/\omega_n^\alpha$ with various values of $\alpha$.
To check the possible influence of the distance function and the initial condition on the solution, we performed several independent calculations for given external parameters with various combinations of $\beta$, $N_\mathrm{max}$, and $\alpha$ from random initial conditions.
Figure~\ref{fig.init_nb5} and \ref{fig.init_nb9} show converged DMFT solutions obtained by two different distance functions ($\alpha=0$ and 1) for $N_b=5$ and $N_b=9$ from various initial conditions. 
Two types of solutions are found in $N_b=5$ case, which are categorized by the number of bath orbitals having positive energy, one and two, from bottom to top.
We also observe solutions with no positive energy bath orbitals if the initial bath parameters are highly biased or the values of $V_l$ are allowed to be very small, but they have much larger spectral gaps and $\chi^2$ in comparison to the other two (not shown).
There are two important considerations in choosing a good initial condition.
One is the distribution of $\epsilon_l$, the energy levels of the bath orbitals in the initial condition.
Because of the singularity of the argument of the distance function at $\omega=0$, the sign of $\epsilon_l$ is hardly changed during the minimization process.
The second factor is the size of $V_l$. An easy example to describe this is the single-site DMFT. The first-order derivative of the distance function with respect to $V_l$ is always zero if $V_l$ is zero.
If one of the $V_l$ is zero in the initial condition or becomes zero in the DMFT iteration accidentally, the corresponding bath orbital cannot contribute the hybridization function.
Then the impurity Hamiltonian works with an effectively smaller number of bath orbitals than the actual $N_b$, although the impurity solver requires computational costs as large as those we need to solve the Hamiltonian with the original $N_b$.
In practice, the two factors are related to each other closely.
The minimization process in the DMFT+ED tends to abandon the corresponding bath orbital by dropping $V$ to zero if the initial location of $\epsilon_l$ is exceedingly far from the range the hybridization function has support.
In order to prevent the bath orbitals from being trapped at $V_l=0$ and to maximize the coverage of given bath orbitals, we apply following procedure in each DMFT iteration.
\begin{itemize}
	\item After new bath parameters are obtained, check the size of $V_l$'s. If any of them is smaller than a threshold value $V_\mathrm{th}$ (here $V_\mathrm{th}=0.01t_{pd}$),
		move the corresponding $\epsilon_l$ to $\epsilon_p+\delta$ ($p=1,2,\cdots,N_b$, $p\neq l$, $V_p \geq V_\mathrm{th}$) with a small random number $\delta$ and set $V_l$ to be finite, here $V_l=0.1t_{pd}$.
		The precise new value of $V_l$ is not important as long as it is non-negligible finite number comparable to $t_{pd}$.
	\item Start the conjugate gradient procedure from the new starting point and obtain a set of new bath parameters.
	\item Choose the set which produces the smallest distance function and resume the DMFT iteration.
\end{itemize}

\begin{figure}[tbp]
	\includegraphics[type=pdf,ext=.pdf,read=.pdf,width=0.99\linewidth]{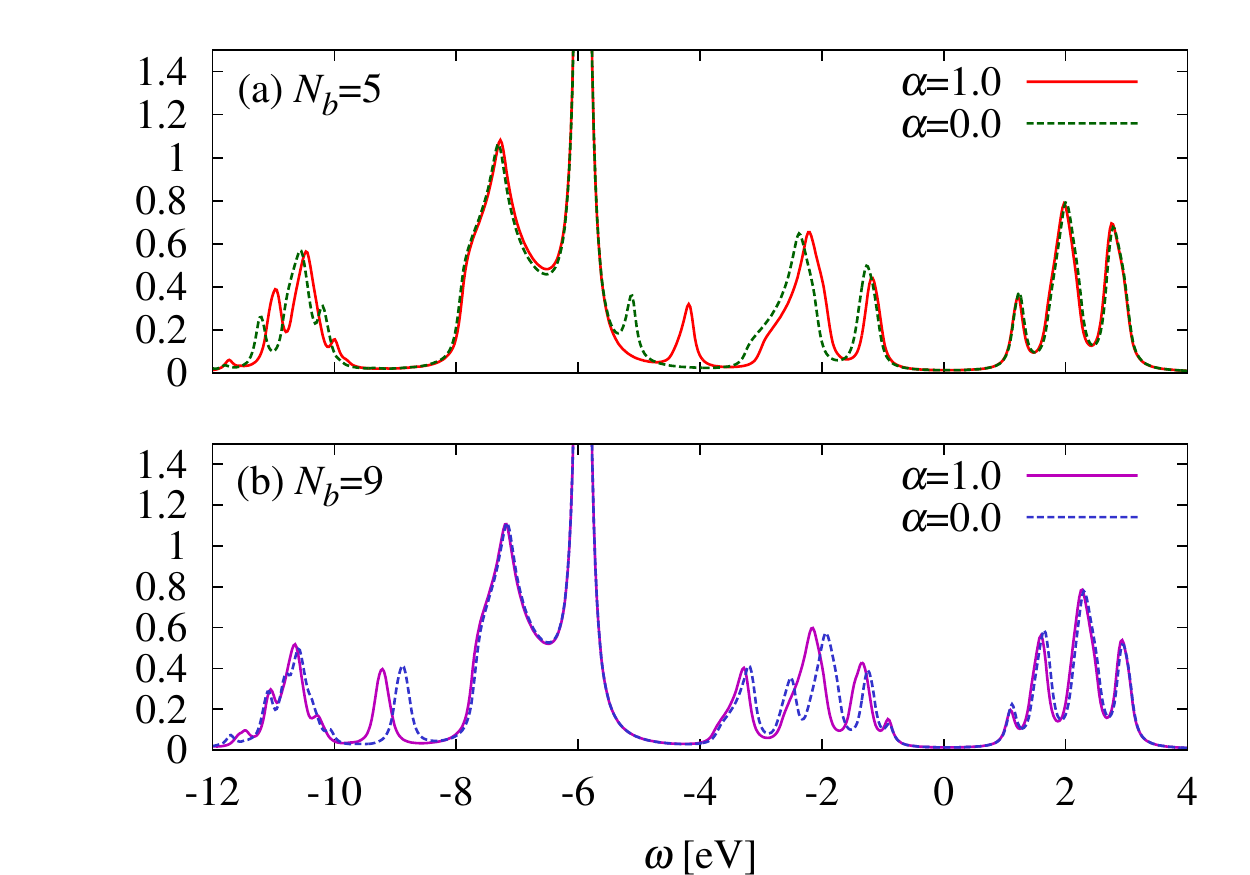}
  \caption{\label{fig.dos_alpha}%
  (color online) Density of states for (a) $N_b=5$ and (b) $N_b=9$. In each combination of ($N_b$, $\alpha$), the solution with smallest $\chi^2$ is presented.
  }
\end{figure}

This eliminates accidentally vanishing hybridizations in the ED+DMFT calculation systematically and allows us to maximize the capability of bath orbitals to mimic the Weiss field.
The rough classification based on the number of positive $\epsilon_l$ is still valid in case of larger $N_b$ as shown in Fig.~\ref{fig.init_nb9}.
Although all the categories are locally stable solutions, one of them ensures smaller distance than the others (The topmost data on Fig.~\ref{fig.init_nb5} and Fig.~\ref{fig.init_nb9}(d) and the second one on Fig.~\ref{fig.init_nb9}(a)).
A larger $N_b$ does not guarantee an easier and less ambiguous minimization. Instead, it develops more locally stable points of the distance function and increases a range of parameters which can be stabilized in the DMFT iteration.
However, use of more bath orbitals enable us to obtain the Weiss field closer to the exact one and reduces the dependence on the details of the distance function: For example, one sees in Fig.~\ref{fig.init_nb9} that the difference between $\alpha=0.0$ and $\alpha=1.0$ solutions is smaller in $N_b=9$ case than $N_b=5$ counterpart.
In other words, even if more stable solutions (or more categories) are found in the DMFT procedure, the cost function dependence within a category is decreased as $N_b$ grows.
We can pick the best category with smallest $\chi^2$ and achieve the best convergence for a given distance function.
There is also minor dependence on $\alpha$ but the difference between the solutions obtained with different $\alpha$ is much smaller than that from the distribution of bath orbitals.
We performed separated calculations with various values of $\beta$ and $N_\mathrm{max}$ (not shown), but the effects are even smaller than $\alpha$-dependence if $\beta>50$ with sufficiently large $\omega_\mathrm{max}$, which is approximately four times of the largest bath energy level.
In the main text, we show results obtained with $\beta=100$, $N_\textrm{max}=400$, $\alpha=0.0$ and $V_\mathrm{th}=0.01t_{pd}$.

\begin{figure}[tbp]
	\includegraphics[type=pdf,ext=.pdf,read=.pdf,width=0.99\linewidth]{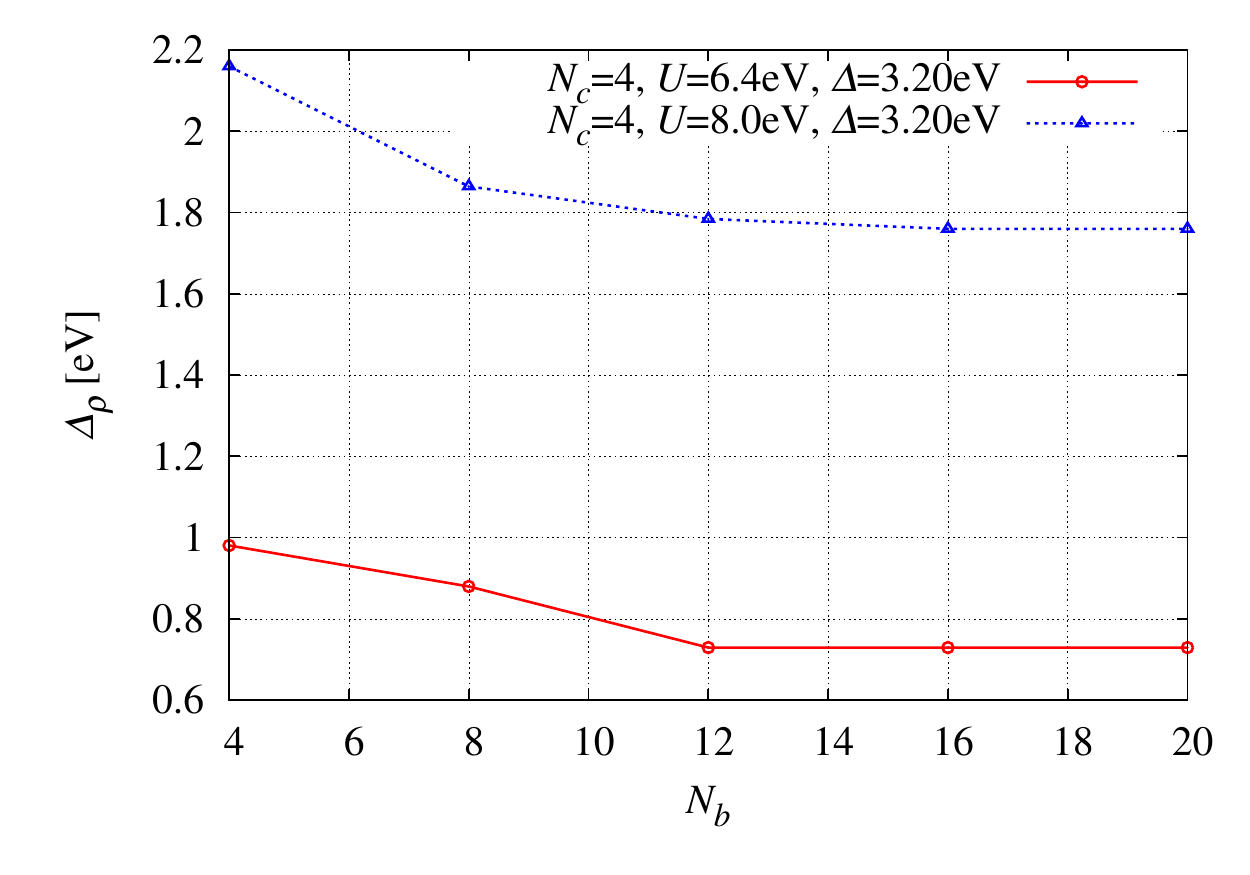}
  \caption{\label{fig.kgap_nb}%
  (color online) $N_b$-dependence of the spectral gap for $N_c=4$, $U=6.4$~eV, and $\Delta=3.20$~eV. $N_b \geq 12$ shows converging behavior.
  }
\end{figure}

In practice, we perform calculations from random initial conditions for a given set of external parameters and choose the solutions give the smallest $\chi^2$.
We can stick to a certain category by putting the initial condition from the chosen solution and by varying the external parameter by small amount.
The starting point can be any point, but a solution with a comparable bath occupancy leads to more rapid convergence to a better fit.
The occupancy is closely related to the distribution of bath energy levels: Larger occupancy corresponds to more bath levels located below Fermi level, i.e., negative energy.
In this target system, for example, the $N_p$ is supposed to be 3-5 (3 for $N_d=2$ and 5 for $N_d=0$). Then roughly 5-9 out of 9 bath orbitals are expected to be filled, or 0-4 orbitals to be empty.
Obviously the optimal value is middle of 0-4 and actual converged solutions are located in that range ash shown in Fig.~\ref{fig.init_nb9}, but we can take advantages of the estimation to set the initial condition at the beginning, before we perform the calculation.

In this section we have described how to find the optimal solution for a give $N_b$.
We perform many independent calculations with various initial conditions and cost functions to pick the best one.
The difference between distinct solutions with a given $N_b$ is considerably smaller than one between solutions from different $N_b$: adding few more bath orbitals improves the fitting much more effectively than tunning the cost function for a small $N_b$.
In this context, the CI solver has great advantages to include more bath orbitals and to obtain better self-consistent solutions of the DMFT equation.

Now the question is how many bath orbitals are required to obtain reliable estimations of the hybridization function for $N_c=4$.
Figure~\ref{fig.kgap_nb} presents the calculated spectral gap from the four-site DMFT as a function of $N_b$.
Although the detailed structure of the DOS is not completely converged, the spectral gap shows converging behavior for $N_b \geq 12$.
The gap decreases continuously as $N_b$ increases, but it approaches a nonzero value.
In the main text, we used $N_b=12$ to compromise between computation costs and numerical accuracy.

\begin{figure}[tbp]
	\includegraphics[width=0.99\linewidth]{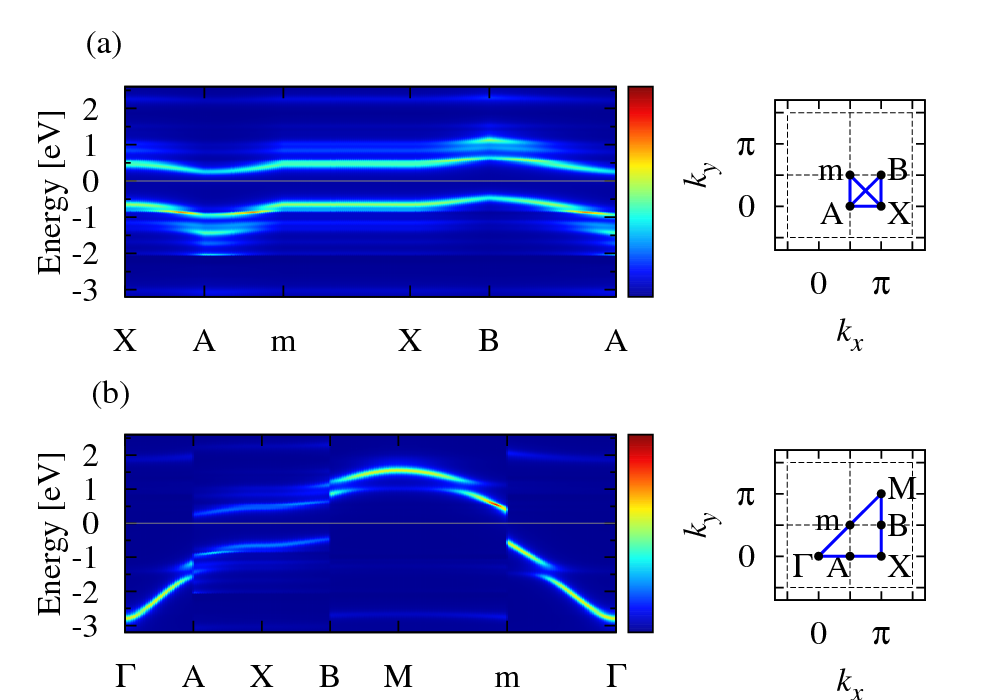}
  \caption{\label{fig.indirect}%
  (color online) Spectral weights on the path given together (a) within $(\pi,0)$-sector and (b) across sectors for $U$=6.4~eV and $\Delta$=3.2~eV.
  The gray solid line marks the Fermi level.
  The highest occupied (lowest unoccupied) level is at B (A) in $(\pi,0)$-sector. 
  The jumps along the path occur when it crosses a boundary between momentum tiles.
  }
\end{figure}

\subsection{Optical conductivity}

At zero temperature, the optical conductivity excluding the vertex correction is calculated as \cite{Wang2011}
\begin{align}
	\sigma_\mathrm{bub}(\Omega)/\sigma_0 =
	\mathrm{Re}\Big[ & \sum_\k \int_{-\Omega}^{0} \frac{d\omega}{2\pi \Omega}
		\Tr \big[
			\mathbf{J}(\k) G(\k, \omega+\Omega+ i\eta) \nonumber \\
			& \times \mathbf{J}(\k) \{G(\k, \omega- i\eta) - G(\k, \omega+i\eta)\}\big]
		\Big],
		\label{eq.bubb}
\end{align}
where $\mathbf{J}$ is the current operator, $\eta$ is a small broadening factor, and $\sigma_0 = e^2/\hbar a_0 \sim 4000 (\Omega \mathrm{cm})^{-1}$ for typical cuprates.
The vertex correction is zero due to the momentum independence of the self-energy in the single-site DMFT, but it is not negligible in the DCA application as Lin \textit{et al.} pointed out \cite{Lin2009}.
In standard tiling with $N_c=4$, the vertex correction solely comes from the jump of the self-energy of the tile boundaries as follows.
\begin{widetext}
\begin{align}
	\sigma_\mathrm{vert}(\Omega)/\sigma_0 =
	 \mathrm{Re}\Big[ \sum_\k \int_{-\Omega}^{0} \frac{d\omega}{2\pi \Omega}
		 \Tr \big[	& \mathbf{J}(\k) G(\k,\omega+\Omega+i\eta) \mathbf{\Gamma}(\omega+\Omega+i\eta, \omega-i\eta) G(\k,\omega-i\eta) \nonumber\\
			 -& \mathbf{J}(\k) G(\k, \omega+\Omega+i\eta) \mathbf{\Gamma}(\omega+\Omega+i\eta, \omega+i\eta) G(\k, \omega+i\eta) \big]
	\Big], \\
	\mathbf{\Gamma}(\omega^\prime, \omega) = & \mathbf{n}^{ab}[\Sigma_b (\omega^\prime) - \Sigma_a(\omega)] \delta \big( (\k-\k^{ab})\cdot \mathbf{n}^{ab} \big),
\end{align}
\end{widetext}
where $\k^{ab}$ is the boundary between momentum tiles a and b, $\mathbf{n}^{ab}$ is a unit normal vector perpendicular to $\k^{ab}$, and $\delta$ is the Dirac delta function.

The spectral weights,
\begin{align}
	A(\k,\omega) = -\frac{1}{\pi}\Tr[ G(\k,\omega+i\eta) ],
\end{align}
enable us to understand the low energy behavior of the optical conductivity. Figure \ref{fig.indirect} illustrates single-particle excitation spectra on the lines given in the figure.
Since the self-energy in DCA is piecewise constant, there are jumps in the spectrum of the four-site DMFT when the path crosses tile boundaries in Fig.~\ref{fig.indirect}(b).
In Fig.~\ref{fig.indirect}, the system has an indirect gap regardless of $N_c$.
The lowest possible optical transition is between A and B, two distinct points in momentum space.
The bubble term, Eq.~(\ref{eq.bubb}), is not allowed to contribute in the transition because it does not carry momentum.
On the other hand, the self-energy in the single-site DMFT is constant, therefore the vertex contribution is not present.
It explains unexpectedly small gap edge conductivity in the single-site DMFT: both terms have zero contribution to the optical conductivity at the gap edge.

\begin{figure}[tbp]
	\includegraphics[width=0.99\linewidth]{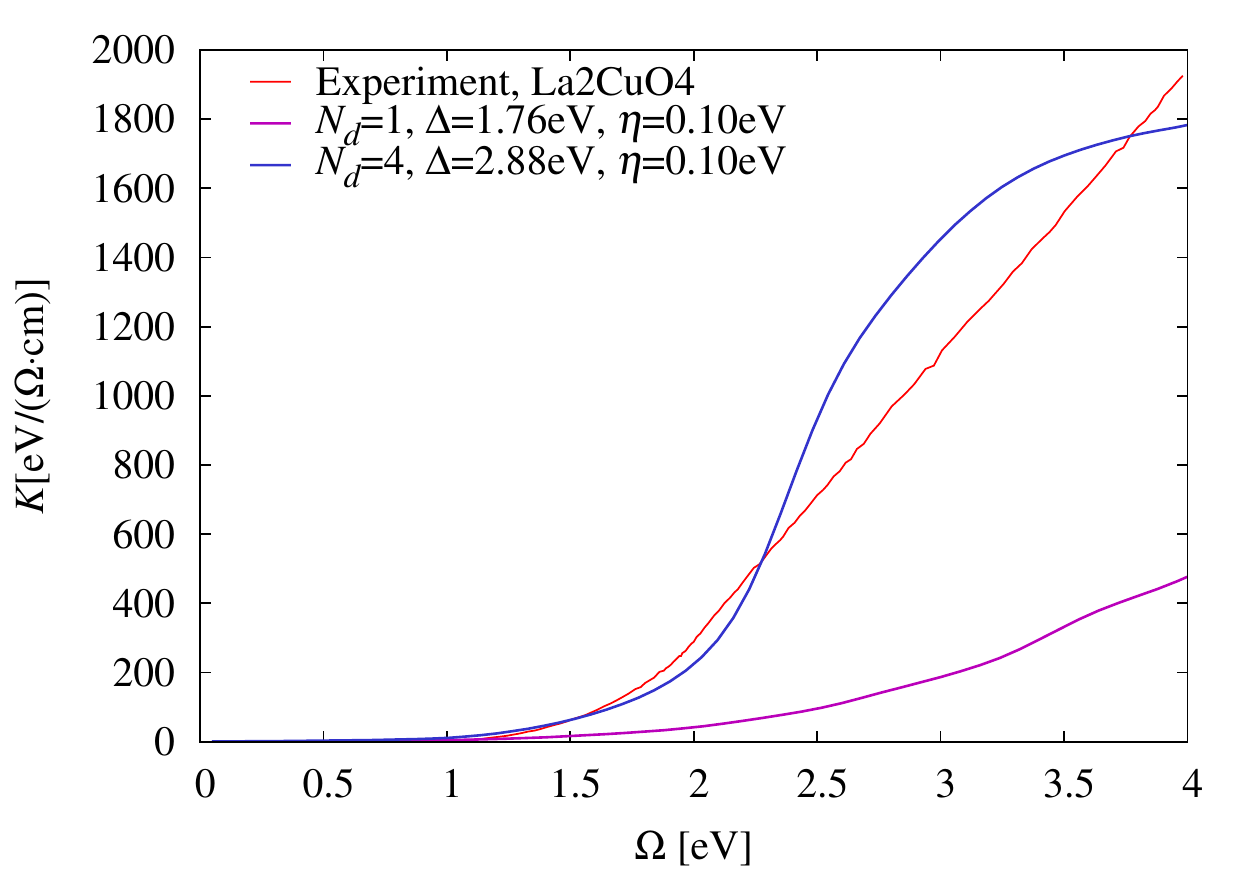}
  \caption{\label{fig.accu}%
  (color online) Integrated optical conductivity for $U$=8~eV and $\Delta$=2.88~eV.
  }
\end{figure}
\begin{figure}[tbp]
	\includegraphics[width=0.99\linewidth]{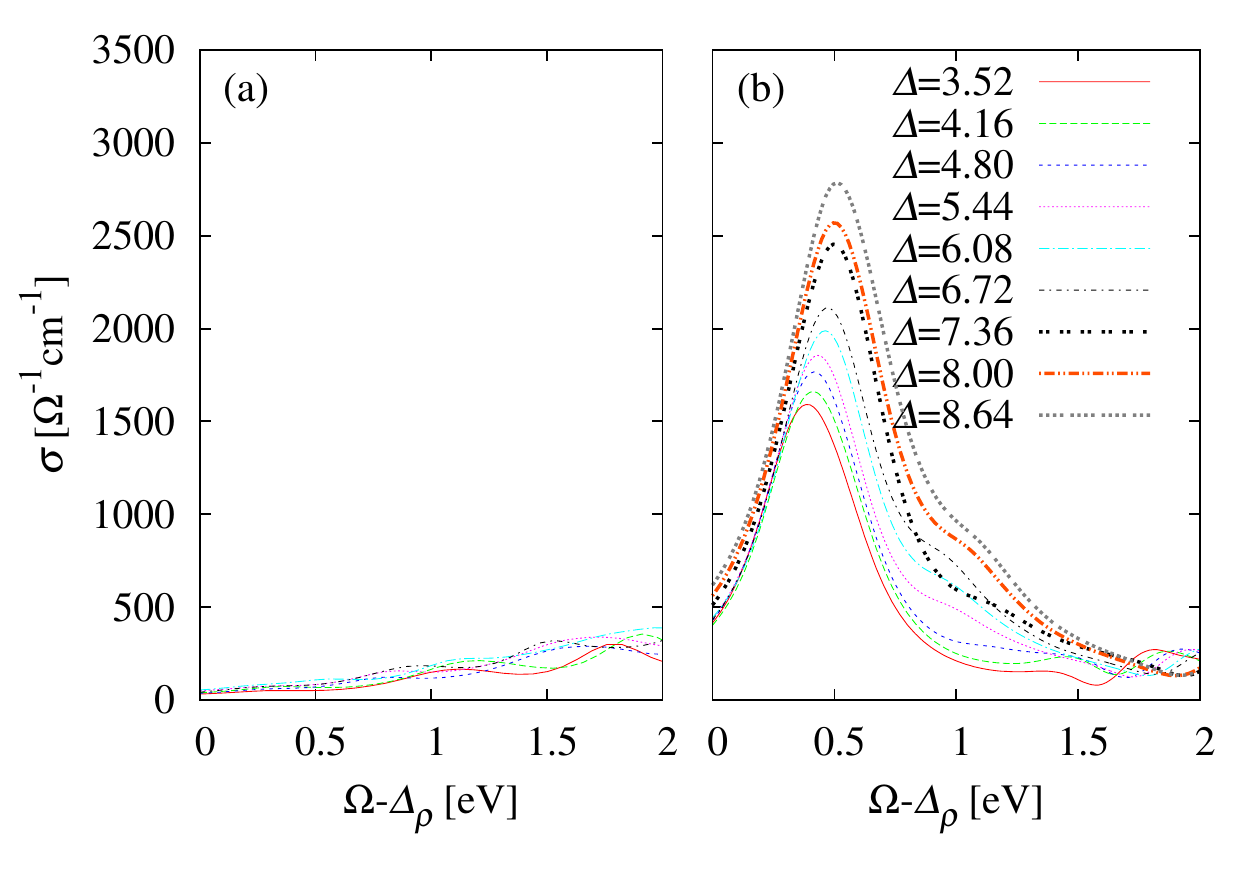}
  \caption{\label{fig.edge}%
  (color online) Gap edge conductivities for $U=11.2$~eV from (a) single and (b) four site DMFT as a function of shifted frequencies by the spectral gap.
  }
\end{figure}

The indirect gap makes the bubble contribution relatively small even in the four-site DMFT, although the tile dependence of the self-energy reduces the direct gap slightly.
However, the large vertex correction develops and it dominates the gap edge conductivity.
In DCA formalism with standard square tiling of momentum space, the vertex function is nonzero only on the boundaries between momentum tiles where the self-energy jumps.
Then the vertex function contributes the conductivity only if the transition is active at the boundaries. It results in delta-peak-like structure at corresponding energy on the conductivity.
In principle, the concentrated weights would be distributed more uniformly if $N_c$ is increased, although it is beyond the computational capacity at this stage.
$N_c=4$ is not sufficient to reproduce the experimental result exactly, however, integrated conductivity $K(\Omega)=\int_0^{\Omega} d\omega \sigma(\omega)$ in Fig.~\ref{fig.accu} shows that four-site DMFT arises significant improvement to explain the experimental result in comparison to the single-site DMFT counterpart which is substantially smaller than expected values.

The vertex correction is closely related to the spatial correlations which is most important in the MCMI phase.
Figure~\ref{fig.edge} presents gap edge conductivities with various values of the charge transfer gap $\Delta$ for both $N_c=1$ and $N_c=4$ cases.
For direct comparison between data of distinct spectral gaps, we shifted frequencies by the spectral gap $\Delta_\rho$, so that the gap edge is located at the origin.
In Fig.~\ref{fig.edge}(a), the single-site DMFT results are not clearly distinguishable at the gap edge.
In spite of the minor enhancement at higher frequencies for larger $\Delta$, the experimental value at gap edge is never reproduced within the single-site DMFT even near the metal-insulator transition.
The four-site DMFT shows completely different behavior.  The conductivity increases as the charge transfer gap grows as shown in Fig.~\ref{fig.edge}(b).
The enhancement becomes larger when the system enters MCMI regime ($\Delta > 6.96$~eV, last three data marked by thicker lines), implying the connection to strong spatial correlations.

\bibliography{cu}
\end{document}